\documentclass[11pt]{article}
\pdfoutput=1
\usepackage{amsmath,amssymb,graphicx,multirow,xspace}
\usepackage[dvipsnames]{xcolor}
\usepackage[colorlinks=true,linktocpage=true]{hyperref}
\hypersetup{                   %
    colorlinks = true,         
    linkcolor  = MidnightBlue, 
    urlcolor   = MidnightBlue,     
    citecolor  = MidnightBlue,  
    anchorcolor = MidnightBlue,
    pagecolor = MidnightBlue
}
\usepackage[sort&compress,numbers]{natbib}
\usepackage{braket}
\usepackage{slashed}
\usepackage{physics}
\usepackage{bm}
\usepackage{indentfirst}
\usepackage{appendix}
\usepackage{verbatim}
\usepackage{cancel}
\usepackage{orcidlink}
\usepackage{titlesec}

\usepackage[T1]{fontenc}
\long\def\symbolfootnote[#1]#2{\begingroup%
\def\thefootnote{\fnsymbol{footnote}}\footnote[#1]{#2}\endgroup}

\input prepictex
\input pictex
\input postpictex
\newdimen\tdim
\tdim=\unitlength

\begin{document}

\begin{titlepage}
\vspace*{-1.5cm}
\hfill{FERMILAB-PUB-25-0853-T}
\vspace{0.5cm}
\begin{center}
{\Large\bf
Widen the Resonance at Ultra-High Energies: \\
Novel Probes of Neutrino Self-Interactions in the High-Mass Regime 
\par}
\end{center}

\vspace{0.2cm}
\begin{center}
{\large
Pedro A. N. Machado~\orcidlink{0000-0002-9118-7354},$^{1}$\symbolfootnote[1]{pmachado@fnal.gov},\,
Isaac R. Wang~\orcidlink{0000-0003-0789-218X},$^1$\symbolfootnote[2]{isaacw@fnal.gov},\, Xun-Jie Xu~\orcidlink{0000-0003-3181-1386},$^{2}$\symbolfootnote[3]{xuxj@ihep.ac.cn},\, Bei Zhou~\orcidlink{0000-0003-1600-8835},$^{1,3}$\symbolfootnote[4]{beizhou@fnal.gov}
}\\
\vspace{0.6cm}
\textit{$\,^1$ Theory Division,
Fermi National Accelerator Laboratory,
Illinois 60510, USA}\\
\vspace{0.5cm}
\textit{$\,^2$ Institute of High Energy Physics, Chinese Academy of Sciences, Beijing 100049, China}\\
\vspace{0.5cm}
\textit{$\,^3$ Kavli Institute for Cosmological Physics, University of Chicago, Chicago, Illinois 60637, USA}
\vspace{0.5cm}
\end{center}

\vspace{0.4cm}

\begin{abstract}
Neutrino self-interactions beyond the Standard Model are well motivated by the nonzero masses of neutrinos, which are the only known particles guaranteed to have new physics. Meanwhile, cosmic messengers, especially neutrinos, play a central role in probing new physics, as they provide experimental conditions far beyond the reach of laboratories and serve as the link between laboratory fundamental-physics discoveries and their roles in the Universe, where many new physics motivations originate.  
In this work, we propose a novel probe of neutrino self-interactions through ultrahigh-energy neutrinos scattering off the cosmic neutrino background when the lightest neutrino species remains relativistic today.
This allows us to ``\textit{Widen the Resonance}'' of such scattering~\cite{Wang:2025qap}.
In addition, we also provide a semi-analytic framework for cosmogenic ultrahigh-energy neutrino production, avoiding computationally intensive simulations and yielding results precise enough for beyond-the-standard-model studies. 
The widened resonance enables future ultrahigh-energy neutrino telescopes, 
in particular GRAND, to probe mediator masses from MeV to GeV, reaching couplings down to $g \sim 10^{-3}$---up to two orders of magnitude beyond current bounds. 
Our results enhance the discovery potential of neutrino self-interactions in the high-mass regime, potentially offering crucial insights into the connections between the neutrino sector and dark sector.
\end{abstract}

\end{titlepage}

\vspace{0.2cm}
\noindent

\noindent\makebox[\linewidth]{\rule{\textwidth}{1pt}}
\tableofcontents
\noindent\makebox[\linewidth]{\rule{\textwidth}{1pt}}

\section{Introduction}

Neutrinos are the only known particles that guarantee connections to new physics, as their nonzero masses are beyond the Standard Model (BSM).
Explaining their masses typically introduces new mediators that naturally lead to BSM neutrino self-interactions ($\nu$SI), which can be much stronger than those in the Standard Model (SM) mediated by $Z$ bosons.\footnote{Hereafter, $\nu$SI solely denotes BSM neutrino self-interactions}
The possible existence of $\nu$SI could also provide a portal to other well-motivated new physics, such as dark matter and matter-antimatter asymmetry. 
Therefore, $\nu$SI provides a promising avenue to discover new physics. 
$\nu$SI have rich phenomena in particle physics, astrophysics, and cosmology. As a result, $\nu$SI has been a subject of growing interest in recent years, using various ways to probe $\nu$SI~\cite{Kreisch:2019yzn,Blinov:2019gcj,Brdar:2020nbj,Deppisch:2020sqh,RoyChoudhury:2020dmd,RoyChoudhury:2022rva,Venzor:2023aka,Das:2017iuj,Shalgar:2019rqe,Chang:2022aas,Fiorillo:2023ytr,Fiorillo:2023cas,Wu:2023twu,Ng:2014pca,Ioka:2014kca,Bustamante:2020mep,Esteban:2021tub,Creque-Sarbinowski:2020qhz,Das:2022xsz,Akita:2022etk,Balantekin:2023jlg,Doring:2023vmk,Luo:2020sho,Huang:2017egl,Chu:2018gxk,Grohs:2020xxd,Li:2023puz,Wang:2023csv,Kaplan:2024ydw,Wang:2025qap,He:2025bex,Leal:2025eou, Poudou:2025qcx,Dev:2024ygx}; see also Ref.~\cite{Berryman:2022hds} for a recent review.

Besides that, cosmic messengers, especially neutrinos, play a central role in probing new physics, providing energies, baselines, and other properties way beyond the reach of laboratory neutrinos (e.g., Refs.~\cite{Ackermann:2022rqc, Wang:2025qap, Leal:2025eou, Bai:2025pef}). 
While laboratory searches for new physics provide a controlled, well-understood setup, many new-physics motivations arise from questions related to cosmology and astrophysics, such as the nature of dark matter.
Thus, cosmic and astrophysical neutrinos provide a natural link between laboratory 
fundamental-physics discoveries and their potential roles in the Universe. 
For $\nu$SI, astrophysical neutrinos provide a powerful test because they can scatter off the cosmic neutrino background (CNB) during propagation through $\nu$SI~\cite{Ng:2014pca,Ioka:2014kca,Esteban:2021tub, Creque-Sarbinowski:2020qhz,Balantekin:2023jlg,Wang:2025qap,He:2025bex,Leal:2025eou}, which would significantly reduce the flux or distort their spectra, both effects being detectable at neutrino telescopes.

In more detail, $\nu$SI are well motivated from the perspective of model building~\cite{Chikashige:1980ui, Gelmini:1980re, Aulakh:1982yn, He:1991qd, Salvioni:2009jp,Lindner:2013awa,Ma:2013yga, Bertuzzo:2017sbj, Babu:2017olk,Berbig:2020wve,Xu:2020qek,Chauhan:2020mgv,Foroughi-Abari:2025upe}. 
They could be mediated by, e.g., a new light scalar or vector particle with mass $m_\phi$ and coupling $g$ to neutrinos. 
The effective four-Fermi interaction strength is $G_X \equiv g^2/m_\phi^2$, currently constrained to $G_X \lesssim 10^2 - 10^9 G_{\rm F}$ from laboratory experiments~\cite{Blinov:2019gcj,Brdar:2020nbj,Deppisch:2020sqh}, supernova dynamics~\cite{Das:2017iuj,Shalgar:2019rqe,Chang:2022aas,Fiorillo:2023ytr,Fiorillo:2023cas}, cosmology~\cite{Huang:2017egl, Chu:2018gxk, Grohs:2020xxd, Li:2023puz}, and astrophysical neutrino propagation~\cite{Ng:2014pca,Ioka:2014kca,Bustamante:2020mep,Esteban:2021tub,Creque-Sarbinowski:2020qhz,Das:2022xsz, Akita:2022etk,Balantekin:2023jlg, Doring:2023vmk}, depending on the mass of the mediator. 
Such new self-interactions could render the Universe partially opaque to ultra-high-energy (UHE) neutrinos through scattering on the CNB.
Of particular importance is resonant scattering\footnote{It is worth mentioning that in the SM, similar resonant scattering could also occur via the $Z$ boson~\cite{Weiler:1982qy}, though this would require much higher energy than what is considered here.} via $\nu  \nu_{\rm CNB} \to \phi \to \nu\nu$, where the $s$-channel mediator production satisfies $s = m_\phi^2$. 
For non-relativistic CNB with neutrino mass $m_\nu \gg T_{\rm CNB} \approx 1.9~{\rm K}=0.16~{\rm meV}$, the resonance condition gives $s \simeq 2 E_\nu m_\nu$, or
\begin{equation}
E_\nu^{\rm res} \simeq \frac{m_\phi^2}{2m_\nu}.
\end{equation}
To probe, e.g., $m_\phi \sim 100$~MeV with $m_\nu \sim 0.05$~eV, requires neutrinos with $E_\nu \sim 100$~PeV, i.e., at high-energy (HE; TeV--PeV) or ultrahigh-energy (UHE; PeV--EeV) energies~\cite{Ackermann:2022rqc}. 
Several studies have investigated this mechanism as a probe of $\nu$SI with IceCube and future IceCube-Gen2 and the Giant Radio Array for Neutrino Detection (GRAND)~\cite{Ioka:2014kca, Ng:2014pca, Kelly:2018tyg, Bustamante:2020mep, Esteban:2021tub, Kelly:2020pcy, Leal:2025eou}.

While the above studies assume a fully non-relativistic CNB, this assumption is not required by neutrino oscillation data, which constrain only $\sum m_\nu \gtrsim 0.06$ eV, allowing one mass eigenstate to satisfy $m_\nu < T_{\rm CNB}$. 
Certain neutrino mass models strictly 
predict a massless neutrino that satisfies this scenario, for example, simple loop-level mechanisms~\cite{Babu:1988ki} or 
minimal
seesaws with only two right-handed standard model singlets~\cite{Abada:2006ea, Abada:2014vea}.
Besides, recent cosmological measurements from DESI suggest $\sum m_\nu$ may be close to the minimum value allowed by oscillations~\cite{DESI:2024mwx}, making a relativistic CNB component increasingly plausible.

For a relativistic neutrino species, the momentum distribution is thermal rather than monochromatic, with energies ranging from $\sim 0.1 T_{\rm CNB}$ to $\sim 10T_{\rm CNB}$.
This thermal distribution significantly alters the resonance phenomenology.
The resonance condition for relativistic scattering is $s = 2 E_\nu E_{\rm CNB}(1-\cos\theta)$, where $\theta$ is the scattering angle and $E_\nu$ is the energy of the incoming UHE neutrino.
Since both $E_{\rm CNB}$ and $\theta$ vary, many different combinations of $(E_\nu, E_{\rm CNB}, 
\cos\theta)$ can satisfy $s = m_\phi^2$. 
This widens the resonance over one order of magnitude in $E_\nu$, in contrast to the narrow absorption feature expected for a non-relativistic CNB. 
This widening was recently demonstrated for the diffuse supernova neutrino background at MeV energies~\cite{Wang:2025qap}, which enables sensitivity to sub-keV mediators and surpasses the existing constraints by orders of magnitude.

The concept of 
``widening the resonance'' is more general than astrophysical neutrinos scattering off relativistic CNB. 
It straightforwardly applies to situations in which both initial states follow continuum energy distributions; for example, other astrophysical neutrinos like the diffuse supernova neutrino background~\cite{Beacom:2010kk} or terrestrial neutrinos.
Thus, we encourage relevant future work in a broader range of topics.

In this work, we extend the 
``widening the resonance'' concept to UHE neutrinos.
The widened resonance provides key advantages: it affects a larger
range of the observable UHE neutrino spectra, yielding greater statistical power and overcoming the potential energy-resolution limitation.
We show that GRAND can probe mediator masses in the MeV-GeV scales with couplings down to $g\sim 10^{-3}$, corresponding to $G_X\sim 10 G_{\rm F}$.
This leads to improvements of up to two orders of magnitude over current constraints, as well as significant gains with respect to the non-relativistic CNB case.
In addition, to model cosmogenic UHE neutrino production, we develop a semi-analytic framework that significantly simplifies computation compared to computationally intensive simulations and can be useful for future studies.

This paper is organized as follows. In Sec.~\ref{sec:model}, we define the general Lagrangian for a scalar-mediated $\nu$SI, with a dedicated description of the flavor structure and the Boltzmann equation that governs neutrino propagation.
In Sec.~\ref{sec:source}, we provide a semi-analytic framework to model the cosmogenic UHE neutrino production.
Then, we illustrate the spectral distortion on UHE neutrinos due to $\nu$SI in Sec.~\ref{sec:propagation}.
The necessary technical details for solving the Boltzmann equations are also discussed.
In Sec.~\ref{sec:detection}, we derive our projected sensitivities from detecting these spectral distortions in GRAND with likelihood analyses. 
In Sec.~\ref{sec:results}, we discuss our results.
We conclude in Sec.~\ref{sec:conclusion}.

\section{Neutrino self-interactions}
\label{sec:model}

In this section, we demonstrate how $\nu$SI attenuates the UHE neutrino propagation.
We begin by setting up the $\nu$SI model considered in this work.
We then illustrate the impact of the relativistic CNB neutrinos by calculating the absorption rate of the propagating UHE neutrinos.
This allows us to estimate the sensitivity based on the mean free path.
Finally, we derive the Boltzmann equation necessary to solve for the UHE neutrino propagation.

\subsection{Lagrangian}

Neutrino self-interactions typically involve a new mediator that couples exclusively or predominantly to neutrinos.
In the simplest setup, the mediator is assumed to be a real scalar, $\phi$, with the following mass and interaction terms:
\begin{align}
\mathcal{L} \supset - \frac{1}{2} m_\phi^2 \phi^2 + \frac{1}{2}( g_{\alpha\beta} \phi \nu_\alpha \nu_\beta + \mathrm{h.c.})\,, \label{eq:L}
\end{align}
where $m_\phi$ denotes the mediator mass, $\nu_{\alpha,\beta}$ with $\alpha,\beta  \in (e, \mu, \tau)$ are SM left-handed neutrinos, and $g_{\alpha\beta}$ is the coupling matrix in the flavor eigenbasis.
Throughout, we assume that neutrinos are Majorana fermions so that right-handed neutrinos are not involved in this work. 
To clarify potential confusions, here we note that $\nu$'s in Eq.~\eqref{eq:L} denote two-component Weyl spinors, instead of four-component Majorana spinors $\chi_M\equiv (\nu, \nu^{\dagger})^T$.
When performing a unitary transformation $\nu\to U \nu$, where $U$ is a unitary matrix, the Majorana spinors cannot transform in the same way as the Weyl spinor unless $U$ is real.  When $U$ is complex, the transformation would be combined with chiral projectors. One could also use Dirac spinors $\psi \equiv (\nu, \eta^{\dagger})^T$ where $\eta$ denotes the Weyl spinor of right-handed neutrinos. In this notation, the interaction term $g_{\alpha\beta} \phi \nu_\alpha \nu_\beta$ becomes $g_{\alpha\beta}\overline{\psi_L^{C}}\psi_L$ where $\psi_L=P_L \psi$ and $\psi_L^{C}$ denotes its charge conjugate. The physical consequences should be independent of whether Dirac, Majorana or Weyl spinors are used.

To calculate the scattering between the propagating UHE neutrino and the CNB neutrino, it is convenient to transform the coupling matrix into the mass eigenbasis.
This is achieved by taking $U$ to be the PMNS matrix, under which the neutrino fields are transformed as:
\begin{equation}
    \left(\begin{array}{c}
            \nu_{e}   \\
            \nu_{\mu} \\
            \nu_{\tau}
        \end{array}\right)=U\left(\begin{array}{c}
            \nu_{1} \\
            \nu_{2} \\
            \nu_{3}
        \end{array}\right),\ \label{eq:PMNS}
\end{equation}
where $\nu_{1}$, $\nu_{2}$, and $\nu_{3}$ are the three neutrino
mass eigenstates.
Correspondingly, the $g_{\alpha\beta}\nu_{\alpha}\nu_{\beta}$
term is transformed to $g_{ij}\nu_{i}\nu_{j}$ with $i,j\in(1,2,3)$.
The transformation of the coupling matrices is given by
\begin{align}
\label{eq:UgU}
    g_{ij} = \sum_{\alpha, \beta} g_{\alpha \beta}U_{\alpha i} U_{\beta j}\,.
\end{align}
Here, both $g_{\alpha \beta}$ and $g_{ij}$ are symmetric matrices.
Throughout this work, we use $g_{ij}$ to represent the coupling matrix in the mass eigenbasis and $g_{\alpha \beta}$ in the flavor eigenbasis.
Note that the transformation in Eq.~(4) is very similar to the transformation of the Majorana mass matrix of neutrinos: $M\to U^TMU$ where $M$ is the mass matrix. Here one should use $U^T$ instead of $U^\dagger$. This is obvious in the Weyl spinor representation. If Dirac or Majorana spinors are used here, one needs to perform the transformation chirally and the same relation can also be obtained.

Since in general $U^{T}U\neq I$ where $I$ is the identity matrix (due to the CP phases in $U$), $g_{ij}$ is not necessarily diagonal even if one assumes a flavor-universal coupling, i.e. $g_{\alpha \beta} = g_{\rm univ} I$.
Specifically, if one neglects the Majorana phases in $U$,
the coupling matrix in the mass eigenbasis can be explicitly calculated as
\begin{equation}
    g_{ij}\approx g\left(\begin{array}{ccc}
            1                         & 0                         & -2ic_{12}s_{13}s_{\delta} \\
            0                         & 1                         & -2is_{12}s_{13}s_{\delta} \\
            -2ic_{12}s_{13}s_{\delta} & -2is_{12}s_{13}s_{\delta} & 1
        \end{array}\right)+{\cal O}\left(s_{13}^{2}\right),\label{eq:g-111}
\end{equation}
where $c_{ij}$ ($s_{ij}$) is $\cos\theta_{ij}$ ($\sin \theta_{ij}$) with $\theta_{ij}$ being the mixing angle between mass eigenstates $i$ and $j$.
The $\delta$ is the complex, CP-violating phase.
One can notice the existence of the off-diagonal elements, though suppressed by the smallness of $s_{13}$.
Taking the limit of $\delta \to 0$ leads Eq.~\eqref{eq:g-111} to $I$.
Throughout this work, we fix the PMNS mixing parameters at their best fit values in NuFIT6.1~\cite{Esteban:2024eli,nu-fit}, 
$\theta_{13}=8.62^{\circ}$, $\theta_{12}=33.76^{\circ}$, $\theta_{23}=43.27^{\circ}$, and $\delta_{{\rm CP}}=207^{\circ}$, 
assuming normal mass ordering.
And for simplicity, we set all Majorana phases to zero.

Before ending this subsection, we comment on possible models that can give rise to the interactions in Eq.~\eqref{eq:L}.
Since $\phi$ is assumed to be exclusively coupled to neutrinos but not charge leptons, one might be concerned about the issue of gauge invariance in model building.
However, given that the electroweak $SU(2)$ symmetry is broken at low energies, a new light mediator arising from BSM theories does not necessarily have to couple equally to neutrinos and their $SU(2)$ partners in their low-energy effective theory.
One of the most well-known examples is the Majoron model~\cite{Chikashige:1980ui, Gelmini:1980re, Aulakh:1982yn}, in which the Majoron is predominantly coupled to neutrinos while its coupling to charged leptons is loop suppressed, roughly by a factor of $G_{\rm F} m_{\nu} m_{\ell}/(16\pi^2)$~\cite{Chikashige:1980ui}.
With separate seesaw mechanisms between different neutrino flavors, a large $\nu_\tau$-philic coupling can be generated~\cite{Blinov:2019gcj}.
Other models that introduce light mediators through the right-handed neutrino portal typically share the same feature~\cite{Berbig:2020wve,Xu:2020qek,Chauhan:2020mgv}.
Therefore, from the model-building point of view, it is plausible that some hidden sector mediators may couple to neutrinos much more strongly than to other SM particles.

\subsection{Resonant absorption and mean-free-path estimate}

In this subsection, we present a qualitative description of the UHE neutrino spectrum distortion by estimating the mean free path of UHE neutrinos, taking into account the dominant effect, i.e., the widened resonance. 
A quantitative study that accounts for all the above effects requires solving the Boltzmann equation, which is presented in the next subsection. 

When a UHE neutrino of energy $E_{\nu}$ propagates through CNB, it could scatter off a CNB neutrino via $\nu  \nu \to \phi \to \nu \nu$.
If the $s$-channel resonance is reached,  $\phi$ can be produced on-shell, absorbing the incoming UHE neutrino.
The cross section at the resonance is
\begin{equation}
\sigma^{(\text{res})}_{ij}\simeq\pi g_{ij}^{2}\delta(s-m_{\phi}^{2})\,,\label{eq:sigma-res}
\end{equation}
where $s=2E_{\nu} E_{\rm CNB} (1-\cos \theta)$ is the Mandelstam variable with $E_{\rm CNB}$ the energy of the CNB neutrino, and $\theta$ the angle between the two initial-state neutrinos.
Later, $\phi$ decays back into two neutrinos, each with roughly half of the energy of the incoming neutrino.
For the relativistic CNB, there is a wide range for  $E_{\rm CNB}$ and $\cos \theta$ to satisfy the resonance condition at $s=m_\phi^2$, allowing for a large range of UHE neutrino energies to resonantly scatter off the CNB neutrinos.
This constitutes the key point of our idea: the 
``widening the resonance'' from scattering the relativistic CNB\footnote{Note that Eq.~\eqref{eq:sigma-res} could also be applied to a non-relativistic CNB, but it is impractical to calculate the $\nu \nu \to \phi$ resonant scattering directly due to its extremely narrow resonance.
Instead, one often incorporates it into the Breit-Wigner formula for the $s$-channel $\nu \nu \to \nu \nu$ scattering---see, e.g., Eq. (25) in Ref.~\cite{Tait:2008zz}.}.

By integrating Eq.~\eqref{eq:sigma-res} with the thermal distribution of CNB neutrinos, we obtain the absorption rate, which is defined as the probability of absorption per unit time (or equivalently, per unit propagation length for UHE neutrinos).
The absorption rate for the $i$-th mass eigenstate is~\cite{Wang:2025qap}
\begin{equation}
\Gamma_{{\rm abs},i}\simeq\frac{g_{i1}^{2}m_{\phi}^{2}T}{16\pi E_{\nu}^{2}}\exp\left[-\frac{m_{\phi}^{2}}{4TE_{\nu}}\right],\label{eq:abs}
\end{equation}
where $T$ is the temperature of the CNB.
Note that within the assumed normal mass ordering only $\nu_1$ contribute to the absorption rate. 
If we had assumed inverted ordering, $\nu_3$ would take that role.

Equation~\eqref{eq:abs} tells us that UHE neutrinos can be effectively absorbed by the CNB during propagation if their energies are in
an appropriate range: 
if 
$E_{\nu}\ll m_{\phi}^{2}/4T$, 
the absorption rate is suppressed by the exponential
factor $\exp(-m_{\phi}^{2}/4TE_{\nu})$; if $E_{\nu}$
is too large, it is suppressed by $E_{\nu}^{2}$ in the denominator. 
The most effective absorptions occur around the
energy that maximizes $\Gamma_{{\rm abs}}$, 
\begin{equation}E_{\nu}^{\text{peak}}=\frac{m_{\phi}^{2}}{8T}\thinspace.\label{eq:E-peak}
\end{equation}
The corresponding absorption rate at this maximum is $\Gamma_{{\rm abs}}=4 g^2T^{3}/(\pi e^{2}m_{\phi}^{2})$.
The smooth exponential form of Eq.~\eqref{eq:abs} leads to a widened resonance peak: the width of the resonance for $E_\nu$ is given by $\Delta E_\nu \sim m_\phi^2/T_{\rm CNB} \simeq 10^9~\rm GeV$ for $m_\phi \simeq 100~\rm MeV$, while the Breit-Wigner resonance width for a non-relativistic CNB is $\Delta E_\nu \sim m_\phi \Gamma_\phi/m_\nu \simeq 1~\rm GeV$ for the same $m_\phi$ and $m_\nu \simeq 0.1~\rm eV$.

Whether our Universe is opaque to UHE neutrinos under $\nu$SI and relativistic CNB can then be estimated by the mean free path at small $z$, with the interaction rate given by Eq.~\eqref{eq:abs}.
For UHE neutrinos moving at the speed
of light, the mean free path is $L=1/\Gamma_{{\rm abs}}$ and the specific value is given as 
\begin{equation}L\simeq3.2~\text{Gpc}\left(\frac{10^{-3}}{|g_{i1}|}\right)^ 2\left(\frac{E_{\nu}}{2\ \text{EeV}}\right)^ 2\left(\frac{50\ \text{MeV}}{m_{\phi}}\right)^{2}e^{\lambda}\thinspace,\label{eq:L-mean}
\end{equation}
where $\lambda\equiv m_{\phi}^{2}/4TE_{\nu}$ is around $1.9$
for $E_{\nu}$ and $m_{\phi}$ varying around their respective benchmark
values in Eq.~\eqref{eq:L-mean}. The mean free path is to be compared with
the radius of the Universe $R_{{\rm Univ}}=14.2\ \text{Gpc}$.
For $E_{\nu}=2\ \text{EeV}$, $m_{\phi}=50\ \text{MeV}$, and $|g_{i1}|=10^{-3}$,
the mean free path $L\simeq21.6$ Gpc is comparable to $R_{{\rm Univ}}$,
implying that the CNB would be half-opaque/half-transparent to UHE
neutrinos at this energy. Consequently, a significant portion of such
UHE neutrinos would be absorbed during propagation, leading to observable
effects in the UHE energy spectrum and flavor composition.
Specifically, when $E_{\nu}$ reaches $E_{\nu}^{\text{peak}}$, the absorption becomes
the most significant and the mean free path $L$ reaches its minimum 
\begin{equation}
    L_{{\rm min}}\equiv\min_{E_{\nu}}\ L\simeq22\ \text{Gpc}\left(\frac{10^{-3}}{g_{i1}}\right)^2\left(\frac{50\ \text{MeV}}{m_{\phi}}\right)^{2}.\label{eq:L-mean-min}
\end{equation}
\begin{figure}
\centering\includegraphics[width=0.7\textwidth]{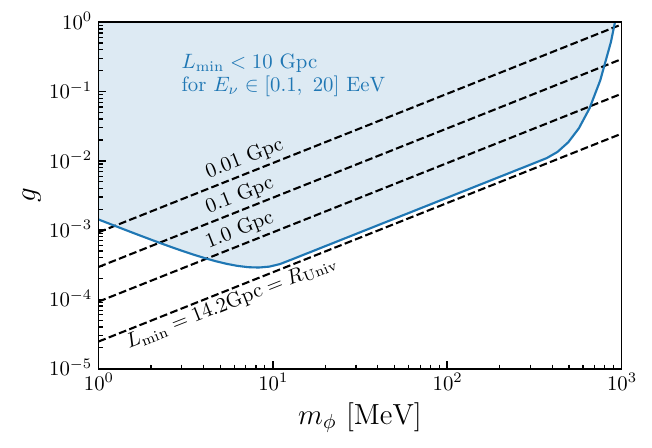}\caption{
Mean-free-path estimate of the $\nu$SI strength
relevant to UHE neutrino propagation. The \textbf{black dashed
lines} are obtained using Eq.~\eqref{eq:L-mean-min}, and the \textbf{{\textcolor[HTML]{2F80B9}{blue
region}}} is obtained by further imposing a finite range of $E_{\nu}$. \label{fig:mean-free-path}}
\end{figure}

Figure \ref{fig:mean-free-path} shows the sensitivity estimated from Eq.\eqref{eq:L-mean} in the blue shaded region. For each point, we scan $E_\nu \in [0.1, 100]$ EeV and take the minimum value of $L$ in this range. The boundary of the shaded region corresponds to $L_{\rm min} \le 10$ Gpc obtained in this way.
For clarity, we also plot several black dashed lines indicating $L_{\rm min} = 0.01$, 0.1, 1.0, and 14.2 Gpc for UHE neutrinos whose energies hit the resonance peak, where $L_{\rm min}$ is computed using Eq.~\eqref{eq:L-mean-min}. Above each line, UHE neutrinos with $E_\nu$ near $E_\nu^{\rm peak}$ undergo strong absorption and are expected to be absorbed within the corresponding interaction length $L_{\rm min}$.
Consequently, if such neutrinos are produced at a distance $D \gg L_{\rm min}$, the flux at $E_\nu = E_\nu^{\rm peak}$ would be exponentially suppressed by a factor $\exp(-D / L_{\rm min})$.

\subsection{Boltzmann equation}

To carefully study the impact of $\nu$SI on UHE neutrino propagation in a quantitative way, we employ the following Boltzmann equation:
\begin{align}
\label{eq:Boltzmann}
    \left(\frac{\partial}{\partial t}-Hp\frac{\partial}{\partial p}\right)F & =-\left[\begin{array}{cc} \Gamma_{\nu(3\times3)}^{-}\\[2mm] & \Gamma_{\phi}^{-}\end{array}\right]F + \left[\begin{array}{c} \Gamma_{\nu(3\times1)}^{+} \\[2mm]\Gamma_{\phi}^{+}\end{array}\right]+\left[\begin{array}{c}S_{\nu(3\times1)} \\[2mm] 0\end{array}\right]\,, \nonumber \\[2mm]
    F & \equiv\left(f_{\nu_{1}}\thinspace,\ f_{\nu_{2}}\thinspace,\ f_{\nu_{3}}\thinspace,\ f_{\phi}\right)^{T}\,,
\end{align}
where $f_{X}$ with $X\in\{\nu_{1},\ \nu_{2},\ \nu_{3},\ \phi\}$
denotes the phase space distribution of particle $X$, $\Gamma_{\nu}^{\pm}$
and $\Gamma_{\phi}^{\pm}$ are production and depletion rates of $\nu$
and $\phi$ via scattering or decay processes, and $S_{\nu}$ is the
astrophysical source term, which will be elaborated in Sec.~\ref{sec:source}.
We use $+$ ($-$) to denote the creation (depletion) rate of a certain particle species.
The subscripts $(3\times3)$ and $(3\times1)$
indicate the shapes of matrices and will be omitted in what follows.
The Hubble parameter $H$ receives contributions mainly from matter
and dark energy for $1\leq1+z\lesssim10$. Hence it can be determined
by
\begin{equation}
    H\approx H_{0}\sqrt{\Omega_{\Lambda}+\Omega_{m}a^{-3}}\thinspace ,\label{eq:H}
\end{equation}
where $H_{0}\approx67.36\ \text{km/s/Mpc}$ is the present value of
$H$, $\Omega_{m}\approx0.315$ and $\Omega_{\Lambda}\approx0.685$
denote the matter and dark energy density parameters~\cite{ParticleDataGroup:2024cfk},
and $a\equiv1/(1+z)$ is the scale factor of the Universe.

The production and depletion rates $\Gamma_{\nu}^{\pm}$ and $\Gamma_{\phi}^{\pm}$,
also known as the collision terms, receive both resonant and non-resonant
contributions.
The resonant contribution ($\propto g_{i1}^{2}$) is
significantly higher than the non-resonant ones $(\propto g_{ij}^{4})$
when the $s$-channel resonance is kinematically accessible.
Under the assumption that the lightest species in the CNB is relativistic,
an UHE astrophysical neutrino, if sufficiently energetic, can always find a relativistic CNB neutrino with appropriate momentum to scatter resonantly.
Therefore, in this work, we only focus on the resonant contribution.
By generalizing the corresponding collision
terms previously derived in Ref.~\cite{Wang:2025qap} to three flavors,
we obtain
\begin{align}
    \Gamma_{\nu}^{-}  & =G_{3\times3}\frac{m_{\phi}^{2}T}{16\pi E_{\nu}^{2}}\exp\left[-\frac{m_{\phi}^{2}}{4TE_{\nu}}\right],\label{eq:-10}                                                                     \\
    \Gamma_{\nu}^{+}  & =G_{3\times1}\frac{2m_{\phi}^{2}}{16\pi E_{\nu}^{2}}\int_{E_{\phi}^{\min}}^{\infty}\dd E_{\phi}f_{\phi}\thinspace,\label{eq:-11}                                                     \\
    \Gamma_{\phi}^{+} & =\sum_i\left(G_{3\times3}\right)_{ii}\frac{m_{\phi}^{2}}{16\pi E_{\phi}p_{\phi}}\int_{E_{\nu}^{-}}^{E_{\nu}^{+}}\dd E_{\nu}f_{\nu_i}\exp\left[-\frac{E_{\phi}-p_{\nu}}{T}\right],\label{eq:-12} \\
    \Gamma_{\phi}^{-} & =G_{1\times1}\frac{m_{\phi}^{2}}{16\pi E_{\phi}}\,,\label{eq:-13}
\end{align}
where $T\approx1.9\ \text{K}\cdot(1+z)$ is the CNB temperature (varying
significantly at high $z$), $E_{X}$ and $p_{X}$ denote the energy
and momentum of $X\in\{\nu,\phi\}$, $E_{\phi}^{\min}=\frac{m_{\phi}^{2}}{4E_{\nu}}+E_{\nu}$,
and $E_{\nu}^{\mp}=(E_{\phi}\mp p_{\phi})/2$.
The matrices $G_{3\times3}$, $G_{3\times1}$, and $G_{1\times1}$ are defined for convenience as follows
\begin{equation}
    G_{3\times3}\equiv\left(\begin{array}{ccc}
            |g_{11}|^{2} & 0            & 0            \\
            0            & |g_{21}|^{2} & 0            \\
            0            & 0            & |g_{31}|^{2}
        \end{array}\right),\ G_{3\times1}\equiv\left(\begin{array}{c}
            \sum_{i}|g_{1i}|^{2} \\
            \sum_{i}|g_{2i}|^{2} \\
            \sum_{i}|g_{3i}|^{2}
        \end{array}\right),\ G_{1\times1}\equiv\sum_{ij}|g_{ij}|^{2}\thinspace.\label{eq:GGG}
\end{equation}
It is noteworthy that if the three components of $\Gamma_{\nu}^{+}$ are summed up, the total creation rate of all neutrino flavors is independent of the PMNS mixing:
\begin{equation}
    \sum_{j}\Gamma_{\nu,j}^{+}\propto\sum_{ij}|g_{ij}|^{2}={\rm Tr}(g_{ij}^\dagger g_{ij})={\rm Tr}(g_{\alpha \beta}^\dagger g_{\alpha \beta}) = \sum_{\alpha\beta}|g_{\alpha \beta}^2|\thinspace.\label{eq:Tr}
\end{equation}

\section{Modeling cosmogenic UHE neutrino production}
\label{sec:source}

In this section, we provide a semi-analytic framework for cosmogenic UHE neutrino production, avoiding computationally intensive simulations and precise enough for phenomenology studies.
The cosmogenic UHE neutrino production
acts as the source term, $\mathcal{S}_\nu$, in Eq.~\eqref{eq:Boltzmann}, which is a crucial ingredient in our work.
We provide a pedagogical description of the semi-analytic approach, which is useful for phenomenological work beyond the scope of this paper.

In general, UHE neutrinos have two main origins (see Refs.~\cite{Ackermann:2022rqc,Muzio:2025xen} for recent reviews). 
The first is cosmogenic, from UHE cosmic rays scattering off the cosmic microwave background (CMB) at temperatures $\sim 10^{-4}$~eV, or the extragalactic background light at $\sim 1$~eV.
The second is directly from astrophysical objects, like certain types of active galactic nuclei or supernovae~\cite{Senno:2015tsn, Chang:2022hqj}.
The neutrino flux from the second origin is highly model-dependent and uncertain, so we ignore this component as a conservative choice. 
Note that if the flux is comparable to or higher than that from the cosmogenic origin, UHE neutrinos will be an even more powerful tool for testing new physics (this is especially motivated by the recent KM3NeT detection of the $\sim200$~PeV neutrino event, which implies a UHE neutrino flux much higher than typical cosmogenic fluxes~\cite{Li:2025tqf, KM3NeT:2025npi}).

The cosmogenic UHE neutrino production can be written as 
\begin{align}
\mathcal{S}_\nu (E_\nu, a) = \frac{2 \pi^2}{E_\nu^2} \frac{\dd N_\nu}{\dd E_\nu} (E_\nu, a) \mathrm{SE}(a) \frac{1}{a^3}\,,
\label{eq:source}
\end{align}
where $a \equiv 1/(1+z)$ is the scale factor and the time variable here.

The source evolution factor $\mathrm{SE}(a)$ in Eq.~\eqref{eq:source} can be parametrized as~\cite{Leal:2025eou,Moller:2018isk}
\begin{align}
\label{eq:SE}
\mathrm{SE}(z) = \begin{cases}
(1+z)^m,                                 & z < 1        \\
2^m,                                     & 1 \leq z < 4 \\
2^m \left( \frac{1+z}{5} \right)^{-3.5}, & 4 \leq z < 7\,.
\end{cases}
\end{align}
The value of $m$ depends on the source type;
for example, redshift evolutions of the star-formation rate, gamma-ray burst, and active galactic nuclei typically follow $m=3$~\cite{Yuksel:2008cu}, whereas BL Lacertae objects
exhibit an overall flat evolution, i.e., $m\simeq 0$~\cite{Ajello:2013lka,Padovani:2015mba,Petropoulou:2016xvm,Qu:2019zln}.

The energy spectra of the produced UHE neutrinos, $\dd N_\nu/\dd E_\nu$ in Eq.~\eqref{eq:source}, is determined by the following processes~\cite{Moller:2018isk, Kelner:2008ke, vanVliet:2019nse, Berezinsky:1969erk, Kotera:2010yn, Aloisio:2015ega, CRPropa:2016qdt}. We ignore the interactions between cosmic rays and the extragalactic background light, as its density is much lower than that of the CMB.
As a simplified benchmark scenario, we treat the UHECRs as all protons, similar to that in Ref.~\cite{Leal:2025eou}.
This assumption might appear too optimistic in light of recent constraints on the chemical composition of UHECRs. 
However, Refs.~\cite{Moller:2018isk,vanVliet:2019nse} have shown that composition and source redshift evolution are degenerate, since for a fixed spectral index, both primarily affect the flux normalization. 
Furthermore, this scenario is not actually the most optimistic for cosmogenic UHE neutrino production, as intermediate-mass nuclei such as He and N can yield even higher fluxes (see, e.g., Appendix A of Ref.~\cite{Moller:2018isk}).

\begin{itemize}

\item {\bf Photopion production:} 
Protons with energies $E_p \gtrsim 55~\rm EeV$ can create $\Delta^+$ resonantly that then decay to a pion and a nucleon, i.e., $p \gamma \to \Delta^+ \to \pi^+ n / \pi^0 p$~\cite{Greisen:1966jv,Zatsepin:1966jv}, with the branching ratio into $\pi^+ n$ being $1/3$. The $\pi^+$ then decays into neutrinos. This is the dominant contribution to the cosmogenic UHE neutrino flux.

\item {\bf Neutron $\beta$ decay:} The neutrons produced together with $\pi^+$ from the above process can decay into $\bar{\nu}_e$. This is suppressed by more than 2 orders of magnitude compared to photopion production for $E_\nu > 0.1~\rm EeV$, because the produced neutrinos have only $10^{-3}$--$10^{-4}$ of the UHECR energies (as a comparison, the photopion produced neutrinos have $\sim 0.05$).
Thus, it is negligible in the energy range considered in this work.

\item \textbf{Multiple scattering:} 
Nucleons produced in the initial interactions can also scatter off CMB photons to generate UHE neutrinos, and the subsequently produced nucleons further contribute in the same way.
However, the steeply falling spectra of UHECRs suppress these additional contributions.

\item {\bf Pair production:} Protons can lose energy due to pair production from scattering off the CMB, i.e., $p \gamma \to p e^- e^+$. The scattering length is $\sim 1$~Gpc, significantly larger than that of photopion production (about 100 Mpc). Consequently, the energy loss from pair production is negligible~\cite{Blumenthal:1970nn,Berezinsky:1988wi,Kelner:2008ke}.

\item {\bf Adiabatic energy loss of UHECRs}: Protons lose energy due to the expansion of the Universe. Note that this corresponds to the interaction length of the parent UHE protons, i.e., between where the protons are born and where they interact and produce UHE neutrinos. 
For the energies considered in our work, the typical interaction length is $\sim 100$ Mpc~\cite{Kotera:2011cp}, thus the protons during this period only lose energy by a factor of 
$\sim 100~{\rm Mpc}/14~{\rm Gpc}\sim 10^{-3}$, which is negligible.

\end{itemize}

Including all the processes above requires computationally intensive simulations~\cite{Moller:2018isk,vanVliet:2019nse, Leal:2025eou}. 
Instead, we provide a simplified, semi-analytic framework for calculating $dN_\nu/dE_\nu$ and show that the results are consistent with those from simulations within theoretical uncertainties, as follows.

\begin{figure}[t!]
\centering\includegraphics[width=0.6\linewidth]{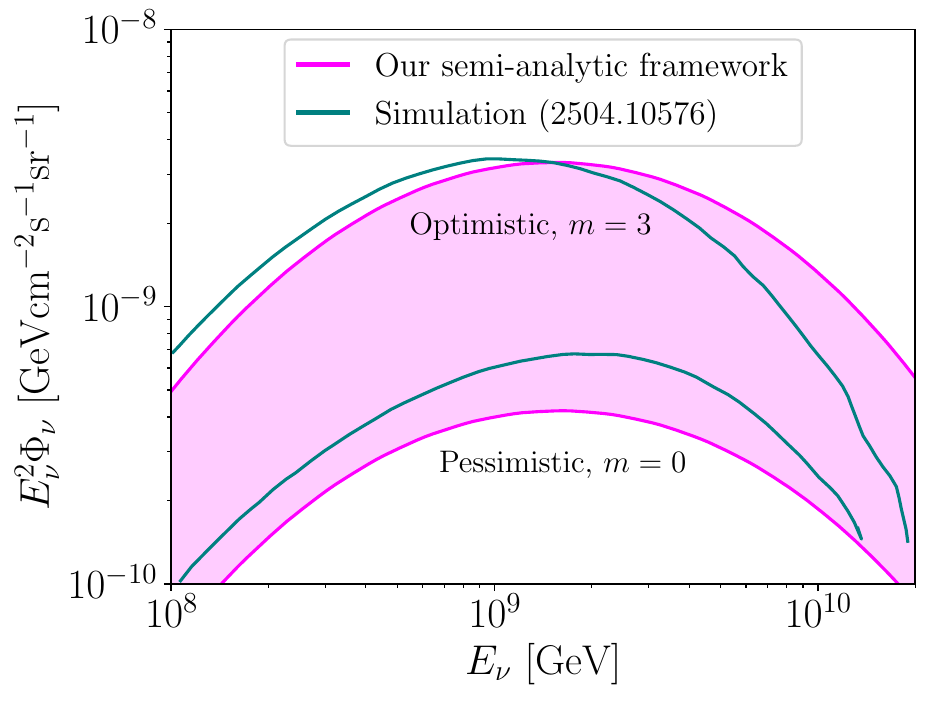}
\caption{
Neutrino energy spectrum (per flavor, e.g., $\nu_\tau + \bar{\nu}_\tau$) without $\nu$SI, for the results from our semi-analytic framework (magenta) and previous simulations (green). 
The agreement between the two calculations is well within the {\it theoretical uncertainty, which spans approximately one order of magnitude} and is illustrated by the shaded band. This band serves as a characterization (rather than a precise quantification) of the theoretical uncertainty. For simplicity, the uncertainty band is shown only for the magenta lines, but a comparable uncertainty also applies to the green lines.
We show the results for our benchmark cases, $\Gamma=2.5$, $E_p^{\rm max}=250$~EeV, with $m=0$ and 3 being optimistic and pessimistic scenarios, respectively; the same conclusion applies to results for other parameter values.
}
\label{fig:source}
\end{figure}

From the above discussion, we focus on the dominant process for cosmogenic UHE neutrino production, i.e., photopion production.
To calculate the neutrino spectra at production, $\dd N_\nu/\dd E_\nu (E_\nu, a)$, of this process, we adopt the widely used semi-analytic framework given in Ref.~\cite{Kelner:2008ke}.
For a given neutrino species $i$,
\begin{align}
\label{eq:dNdEnu}
\frac{\dd N_{\nu, i}}{\dd E_\nu}(E_\nu, z) = \int f_p(E_p) f_{\rm CMB}(\epsilon, z) \Psi_i(\eta, x) \frac{\dd E_p}{E_p} \dd \epsilon\,,
\end{align}
where $x \equiv E_{\nu, i}/E_p$ is the energy ratio between the produced neutrino and the proton, $\eta \equiv 4 \epsilon E_p/m_p^2$ 
is the center of mass (CoM) energy of the proton-photon scattering, normalized by $m_p$.
The integration is over the proton energy $E_p$ and the CMB photon energy $\epsilon$.
The crucial quantity is $\Psi(\eta, x)$, which is the neutrino spectrum per proton-photon interaction.
We refer the readers to Appendix~\ref{app:source} for the details about the form of $\Psi$ and the range of $\eta$.
This quantity is convoluted with the CMB photon spectrum at redshift $z$, $f_{\rm CMB}(\epsilon, z)$, and the UHE cosmic-ray proton energy density, $f_p(E_p)$, which is directly related to their spectrum~\cite{PierreAuger:2020qqz, Fenu:2023men}
\begin{align}
\label{eq:proton power}
\frac{\dd N_p}{\dd E_p} = N_p E_p^{-\Gamma} \exp \left( - \frac{E_p}{E_p^{\rm max}} \right)\,,
\end{align}
where $N_p$ is the normalization factor to match the local ($z \simeq 0$) measurements~\cite{PierreAuger:2020qqz, Fenu:2023men} around 40 EeV~\cite{AlvesBatista:2019rhs, Leal:2025eou}, as UHECRs can only travel distances of $\sim100$~Mpc.
In addition, $\Gamma \in [2, 3]$ is the spectral index and $E_p^{\rm max}$ is the cut-off energy.

With these fitted functions, one can perform a numerical integration for Eq.~\eqref{eq:dNdEnu}.
The results, together with 
the source evolution factor in 
Eq.~\eqref{eq:SE}, are then used to calculate the neutrino production in
Eq.~\eqref{eq:source}.
Finally, because the neutrino oscillation length is much smaller than any astrophysical scales here and the neutrino mixing angles are large, we expect an approximately flavor-independent neutrino flux soon after production, i.e., all flavors have the same flux. 
More concretely, the flavor compositions at detection ($R_{\oplus}$)
and at the source ($R_{\star}$) are related by $R_{\oplus}=\mathbb{P}R_{\star}$
with $\mathbb{P}$ given by~\cite{Xu:2014via}
\begin{equation}
\mathbb{P}=\left(\begin{array}{ccc}
1-Y-Z & Z & Y\\
Z & 1-X-Z & X\\
Y & X & 1-X-Y
\end{array}\right),\label{eq:PP}
\end{equation}
where $(X,Y,Z)\equiv\sum_{i}(\left|U_{\mu i}U_{\tau i}\right|^{2},\left|U_{\tau i}U_{ei}\right|^{2},\left|U_{ei}U_{\mu i}\right|^{2})$.
If one takes the tri-bimaximal mixing pattern ($s_{13}=0$, $s_{23}=1/2$,
$s_{12}=1/3$) that is often used as a simple approximation of the
PMNS matrix, one obtains $(X,Y,Z)=(7/18,2/9,2/9)$. If the production
is dominated by $\pi^{\pm}$ decay, which gives $R_{\star}=(1,\ 2,\ 0)^{T}$,
the $\mathbb{P}$ matrix with the tri-bimaximal mixing converts it
to $R_{\oplus}=(1,\ 1,\ 1)^{T}$. If the best-fit values are used
($\theta_{13}=8.62^{\circ}$, $\theta_{12}=33.76^{\circ}$, $\theta_{23}=43.27^{\circ}$,
and $\delta_{{\rm CP}}=207^{\circ}$), the resulting $R_{\oplus}$
becomes $(0.98,\ 1.03,\ 0.99)^{T}$, which only slightly deviates
from the tri-bimaximal result. The small deviation is negligible in this work, given the precision of UHE neutrino observations. 
Note that for neutrinos from beta decay, which has $R_{\star}=(1,\ 0,\ 0)^{T}$, the corresponding $R_{\oplus}$ becomes $(0.55,\ 0.21,\ 0.24)^{T}$, significantly different from $(1,\ 1,\ 1)^{T}$. However, as mentioned above, the contribution of beta decay at such high energies to the cosmogenic neutrino flux is subdominant.

Fig.~\ref{fig:source} shows the neutrino energy spectrum without $\nu$SI calculated by our semi-analytic framework and, as a comparison, the result from simulations~\cite{Leal:2025eou}.
The difference between the two calculations is much smaller than the overall theoretical uncertainty, which spans approximately an order of magnitude. This validates our semi-analytic framework.

\section{Neutrino propagation}
\label{sec:propagation}

With the collision and source terms constructed, we can solve the Boltzmann equation numerically. 
Although in the original Boltzmann equation, $F$ is a function of $t$ and $p$, in practice, we solve the equation using the variable transformation from $(t,\ p)$ to
$(a,\ \tilde{p})$ where $\tilde{p}\equiv ap$ is the comoving momentum.
Such a variable transformation not only removes the $Hp\partial_{p}$
term in Eq.~\eqref{eq:Boltzmann}, but also improves the numerical
stability and performance of the code, since the comoving momentum is unchanged during propagation in the expanding
Universe in the absence of collisions.
For the technical details of the implementation, we refer to the appendix of Ref.~\cite{Wang:2025qap}.

By discretizing the comoving momentum space into 200 bins for each
particle species, we solve the $200\times4$ coupled differential
equations using the {\tt solve\_ivp} solver in {\tt scipy} with
the {\tt BDF} method.
From the obtained values of $f_{\nu,i}$,
it is straightforward to calculate the present-day UHE neutrino flux
$\Phi_{\nu,i}$, which is related to $f_{\nu,i}$ by
\begin{equation}
    \Phi_{\nu,i}=\frac{E_{\nu}^{2}}{2\pi^{2}}f_{\nu,i}\thinspace,\ \ (i=1,\ 2,\ 3)\thinspace.\label{eq:Phi-i}
\end{equation}
Note that during propagation, the UHE neutrino fluxes are calculated in the mass basis, while their detection involves fluxes in the flavor
basis, $\Phi_{\nu,\alpha}$, which are related to $\Phi_{\nu,i}$ by
\begin{equation}
    \Phi_{\nu,\alpha}=\sum_{i}|U_{\alpha i}|^{2}\Phi_{\nu,i}\thinspace.\label{eq:Phi-alpha}
\end{equation}

We investigate the following two representative scenarios of $\nu$SI
\begin{itemize}
    \item Flavor-universal coupling:  $g_{\alpha \beta}=g_{\rm univ} I$, where $I$ is the identity matrix and $g_{\rm univ}$ denotes the coupling strength.
          This scenario leads to an approximately diagonal coupling matrix in
          the mass basis---see Eq.~\eqref{eq:g-111}.
    \item $\nu_{\tau}$-philic coupling: $g_{\alpha \beta}={\rm diag}\left(0,0,g_{\tau \tau}\right)$.
          This scenario is the least constrained by laboratory bounds~\cite{Blinov:2019gcj}, while we specifically emphasize the importance of future sensitivity explored in this work.
\end{itemize}

\begin{figure}[th!]
    \centering
    \includegraphics[width=0.49\linewidth]{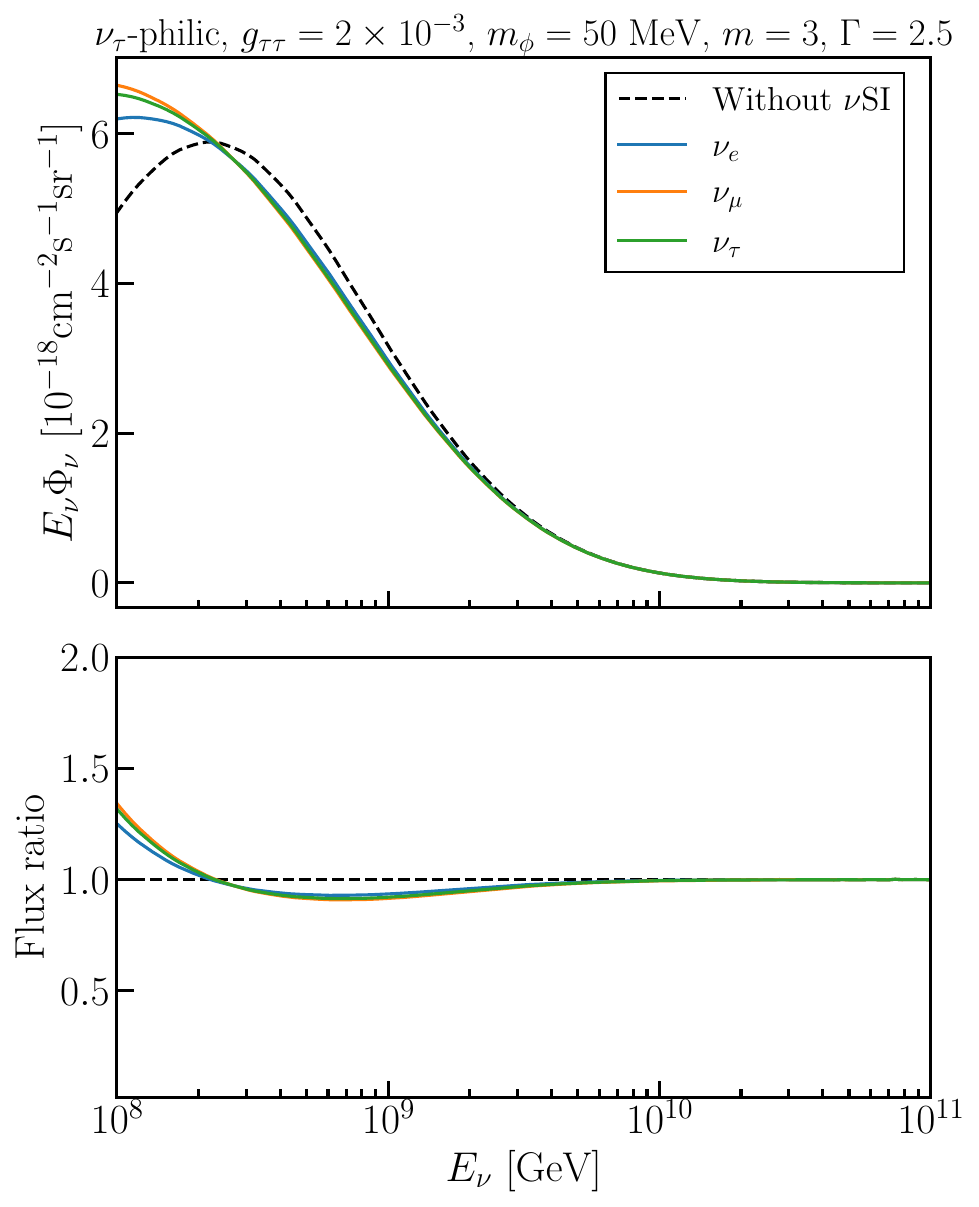}
    \includegraphics[width=0.49\linewidth]{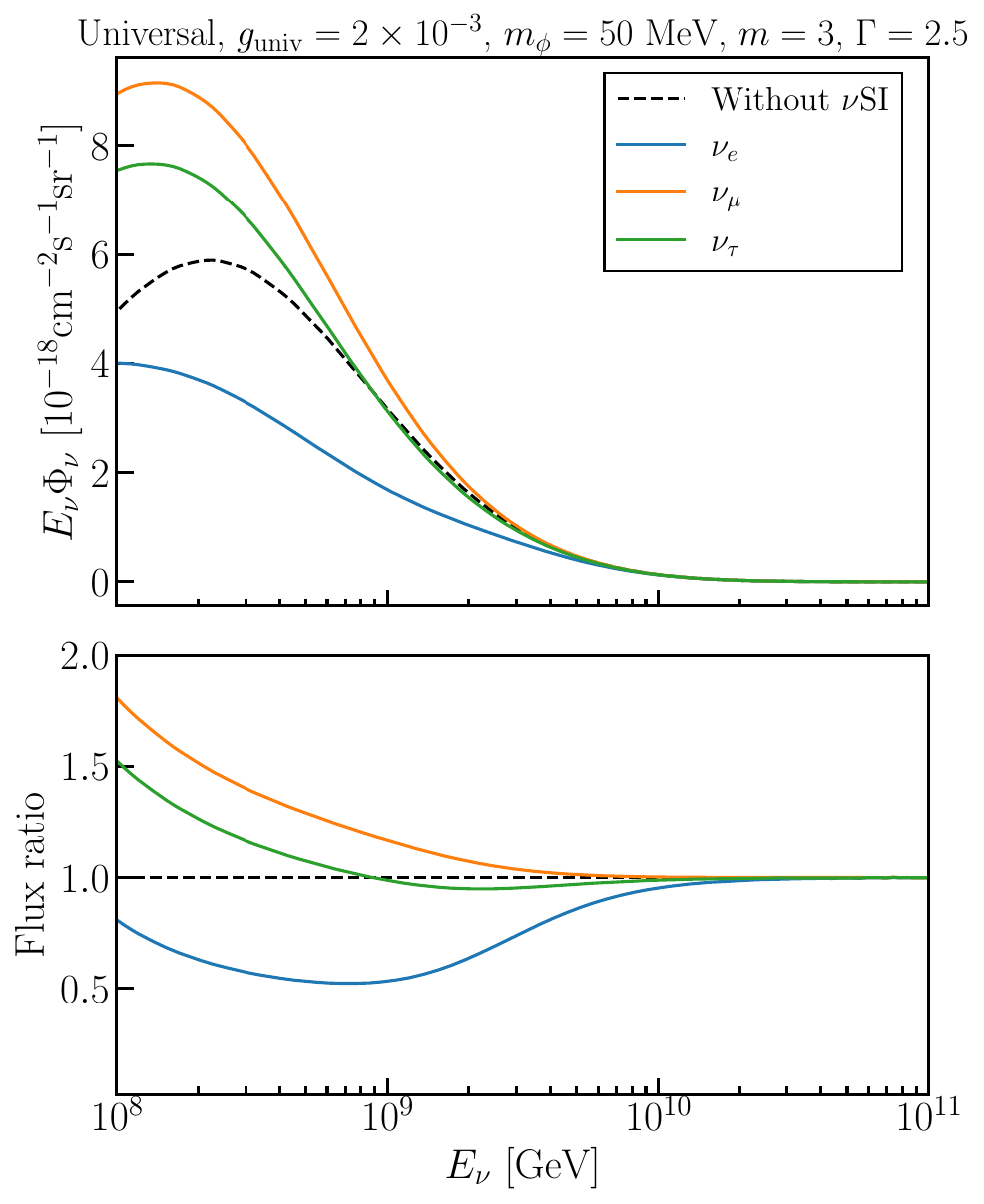}
    \caption{
        UHE neutrino fluxes modified by $\nu$SI. The left and right panels assume flavor universal and $\nu_{\tau}$-philic couplings, respectively.
        The upper panels present the flux spectrum, whereas the lower panels show the ratio of the modified flux to the standard one with free propagation.
}
    \label{fig:spectrum}
\end{figure}

In Fig.~\ref{fig:spectrum}, we present the UHE neutrino spectra obtained
by solving the Boltzmann equation for the two scenarios defined above.
The black dashed curves denote free propagation (with the
source parameters $E_{p}^{\rm max} = 250~\rm EeV$, $\Gamma=2.5$, and $m=3$) while the solid curves include the effect of $\nu$SI with $g_{\tau \tau}  (g_{\rm univ})=2\times10^{-3}$, $m_{\phi}=50~{\rm MeV}$.
In general, all spectra exhibit absorption at high energies and regeneration at low energies. 
In the $\nu_{\tau}$-philic scenario (the left panel), these effects are similar for all three flavors. In contrast, in the flavor-universal scenario (the right panel), $\nu_e$ shows a more prominent absorption feature, while $\nu_{\mu, \tau}$ spectra display stronger regenerations.
In both scenarios, the absorption feature appears as a smooth, widened dip rather than a deep, narrow resonance peak, as a result of the widened resonance.

The key difference arises from the nearly diagonal structure of the coupling matrix in the mass basis for the universal coupling case---see Eq.~\eqref{eq:g-111}.
In the limit of $s_{13}\to0$, Eq.~\eqref{eq:g-111} becomes exactly diagonal, implying that only $\nu_{1}$ can be absorbed by the relativistic CNB.
If all $\nu_{1}$ particles are fully absorbed without regeneration of new
particles, the resulting UHE flux would contain only $\nu_{2}$
and $\nu_{3}$, leading to the following flavor composition: 
\begin{equation}
    \Phi_{\nu,e}:\Phi_{\nu,\mu}:\Phi_{\nu,\tau}\approx s_{12}^{2}:1-c_{23}^{2}s_{12}^{2}:1-s_{23}^{2}s_{12}^{2}\approx \frac13\,:\,\frac56\,:\,\frac56\thinspace,\label{eq:3f-ratio}
\end{equation}
since $s_{12}^{2}\approx1/3$ and $c_{23}^{2}\approx s_{23}^{2}\approx1/2$. This ratio shows that the $\nu_{e}$ flux is strongly suppressed while the reductions in $\Phi_{\nu,\mu}$ and $\Phi_{\nu,\tau}$
are relatively modest, assuming no neutrino regeneration and a perfectly diagonal  $g_{ij}$.
We then include the effect of neutrino regeneration from $\phi$ decay.
Since the decay rate into each flavor is the same, the regenerated UHE neutrinos compensate all three flavor fluxes equally (though at lower energies).
Combining both absorption and regeneration effects, the $\nu_e$ flux remains dominated by absorption, while the final $\nu_\mu$ and $\nu_\tau$ fluxes become larger than their original values due to the regenerated neutrinos produced from absorbed $\nu_e$.
In comparison, in the $\nu_{\tau}$-philic scenario, the absorption effect is much more evenly distributed among $\nu_{1}$, $\nu_{2}$, and $\nu_{3}$.
Using the latest global fit central values of the PMNS mixing parameters, we obtain
\begin{equation}
    g_{ij}\approx g_{\tau \tau}\left(\begin{array}{ccc}
            0.25 & 0.29 & 0.32 \\
            0.29 & 0.33 & 0.37 \\
            0.32 & 0.37 & 0.42
        \end{array}\right),\label{eq:Gmass-tau}
\end{equation}
which shows that the absorption effect on $\nu_{3}$ is only slightly
stronger than on $\nu_{2}$ and $\nu_{1}$.
Neglecting this small difference, all three mass eigenstates experience nearly the same degree of absorption.
As a result, the three curves for $\nu_{e}$, $\nu_{\mu}$, and $\nu_{\tau}$ in the right panel of Fig.~\ref{fig:spectrum} lie very close to one another.

\section{Ultrahigh-energy neutrino detection and likelihood analysis}
\label{sec:detection}

\begin{figure}[h!]
\includegraphics[width=0.47\textwidth]{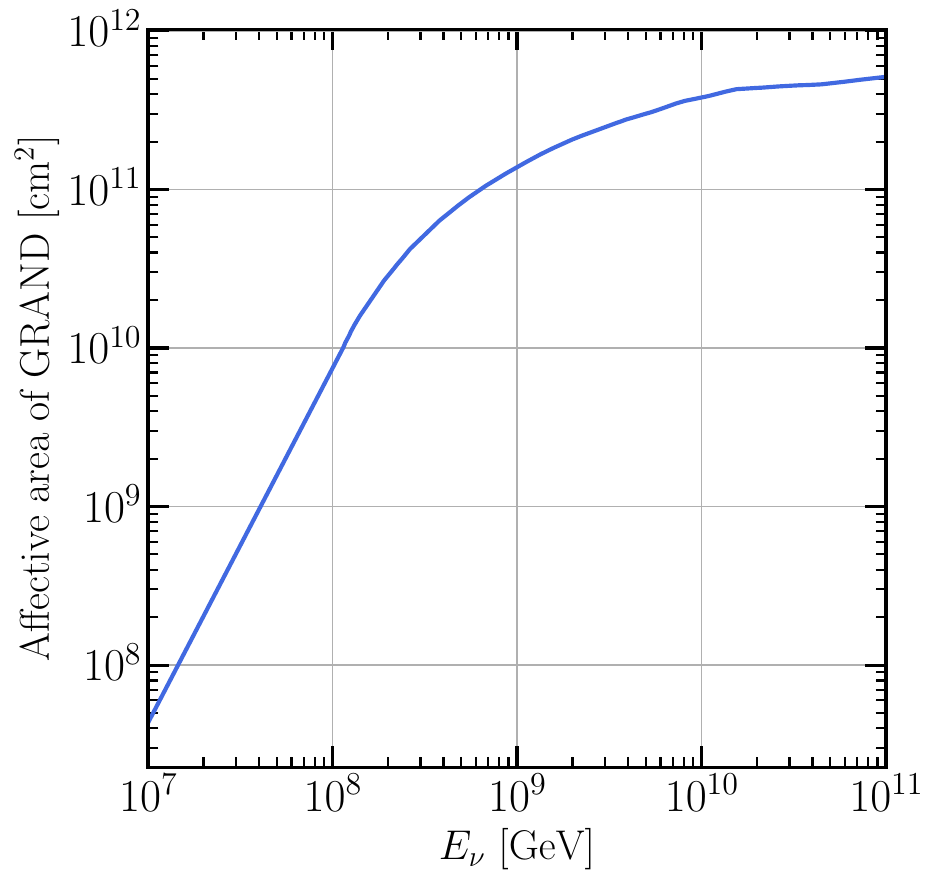}
\includegraphics[width=0.462\textwidth]{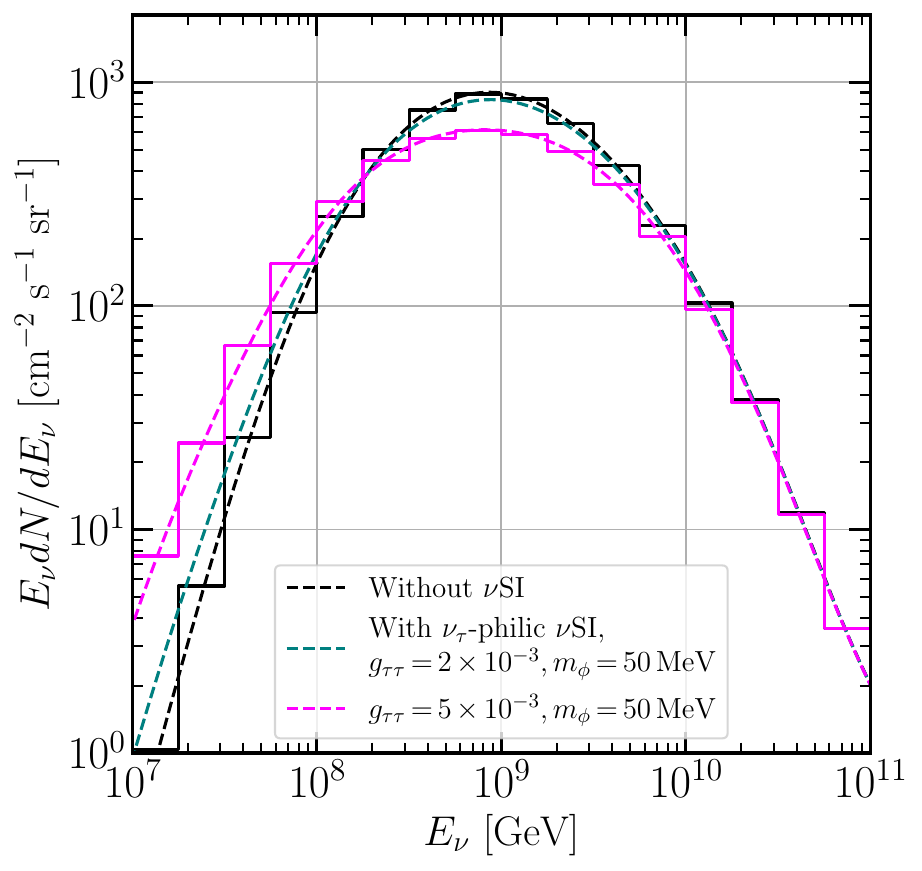}
\caption{
\textbf{Left}: direction-averaged effective area of GRAND for UHE tau neutrino detection. 
\textbf{Right}: our calculated event rates of UHE neutrinos at GRAND with ten years of exposure. 
We also use two histograms to show our energy binning.
}
\label{fig: detection}
\end{figure}

For the UHE neutrino detection, we focus here on the Giant Radio Array for Neutrino Detection
(GRAND)~\cite{GRAND:2018iaj}. GRAND is one of the most representative experiments among those employing a promising experimental strategy that relies on the observation of extensive air showers
(EAS) from the decay of Earth-emergent tau leptons induced by tau-neutrino charged-current interactions~\cite{Ackermann:2022rqc}. 
To observe the EAS, GRAND aims to measure their radio emission
 by deploying a large-scale
network of up to 200,000 autonomous radio antennas over an area of
about 200,000 $\text{km}^{2}$, primarily in mountainous regions. 
We focus on UHE tau neutrinos as the other flavors ($\nu_{e}$ and $\nu_{\mu}$) can hardly
produce ``Earth-emergent'' air showers suitable for observations.

With the direction-averaged effective area, $A_{\rm eff}$, and $T_{\rm obs} = 10$~years of exposure, it is straightforward to calculate the event
rate by
\begin{equation}
\frac{\dd N}{\dd E_{\nu}}
=
4\pi \, T_{\rm obs} A_{{\rm eff}}(E_\nu) \Phi_{\nu,\tau}(E_\nu) \,.
\label{eq:event-rate}
\end{equation}

Figure~\ref{fig: detection} left and right panels show the $A_{\rm eff}$~\cite{GRAND:2018iaj} and our calculated event rates, respectively. 
Within ten years, GRAND is expected
to detect hundreds of tau neutrinos above 1 EeV.
The spectral distortion across a broad energy range due to relativistic CNB yields greater statistical power and overcomes the potential energy-resolution limitation.
Note that the $A_{\rm eff}$ should be corrected due to certain SM processes, like the neutrino-nucleus W-boson production~\cite{Seckel:1997kk,Alikhanov:2015kla, Zhou:2019frk,Zhou:2019vxt,Xie:2023qbn} and final-state radiation~\cite{Plestid:2024bva}, the former increases the $A_{\rm eff}$ by overall $\sim10$\% while the latter deceases $A_{\rm eff}$ by overall $\sim15$\%. However, both effects are much smaller than current uncertainties in the UHE neutrino production so we leave these corrections for future studies\footnote{Note that for another promising UHE neutrino detection strategy, in-ice radio detectors, final state radiation increases the $A_{\rm eff}$ by as much as $\sim60$\%~\cite{Plestid:2024bva}, which should be included in the phenomenology studies~\cite{Plestid:2024bva}.}

Next, to evaluate the sensitivity of UHE neutrino observation at
GRAND to $\nu$SI, we perform a likelihood analysis similar to Ref.~\cite{Leal:2025eou}. The likelihood function is given by~\cite{ParticleDataGroup:2024cfk}
\begin{equation}
\chi^{2}(m, \Gamma, E_p^{\max} | g, m_\phi) 
= 2\sum_{k}\left[N_{\text{st,}k}-N_{k}+N_{k}\ln\frac{N_{k}}{N_{\text{st,}k}}\right],\label{eq:chi}
\end{equation}
where $N_{\text{st,}k}$ and $N_{k}$ denote the event numbers in
the $k$-th energy bin in the without and with $\nu$SI, respectively.
In our analysis, we marginalize over 
$m \in [-3.0, 3.0]$,
$\Gamma \in [2.0, 3.0]$,
and $E_p^{\max} \in [10^2, 10^5]$~{\color{blue}GeV} around two benchmark points,
optimistic scenario ($m=3$) and
pessimistic scenario ($m=0$), both with $\Gamma=2.5$, $E_{p}^{\rm max}=250$~EeV; this encompasses the uncertainties in the cosmogenic UHE neutrino production.
We include the parameters $\lambda$ and $m$ to account for astrophysical
uncertainties
in $N_{\text{st,}k}$ and marginalize them in the frequentist treatment~\cite{ParticleDataGroup:2024cfk}.
The energy bins are from 0.01
to 100 EeV 
with logarithmic bin widths $\log_{10}(E_{\rm high}/E_{\rm low})=0.25$,
accounting for the energy resolution of GRAND; thus we have 16 bins in total.

Before concluding this section, we briefly clarify the motivation for our choice of parameter ranges in the scan, as well as the impact of astrophysical uncertainties, in particular the interplay between UHECR composition and source evolution.
Current measurements of the UHECR composition remain uncertain. While Auger data suggest a trend toward heavier composition at the highest energies, the interpretation is model dependent~\cite{PierreAuger:2024flk}, and Telescope Array results remain compatible with lighter, proton-dominated scenarios~\cite{TelescopeArray:2026rdu}. Moreover, these measurements probe primarily the local Universe ($z \lesssim 0.02$), whereas cosmogenic neutrinos are sensitive to the full cosmological source distribution.
Moreover, a key point is the strong degeneracy between composition and source evolution for the cosmogenic neutrino production: heavier compositions suppress the neutrino flux in a way qualitatively similar to reducing the source evolution parameter $m$. From an astrophysical perspective, most candidate source classes (e.g., star formation, supernovae, gamma-ray bursts, and active galactic nuclei) are expected to exhibit positive evolution, i.e., $m>0$. 
As a result, scanning over $m$ from $-3$ to $3$ within proton-dominated scenarios spans a broad range of physically relevant cosmogenic neutrino fluxes. Finally, we emphasize that our choices of parameter ranges are consistent with previous studies of $\nu$SI with cosmogenic UHE neutrinos~\cite{Leal:2025eou}, which enables a direct comparison between their conventional and our relativistic $\nu$SI scenarios.

\begin{figure}[t!]\includegraphics[width=0.49\textwidth]{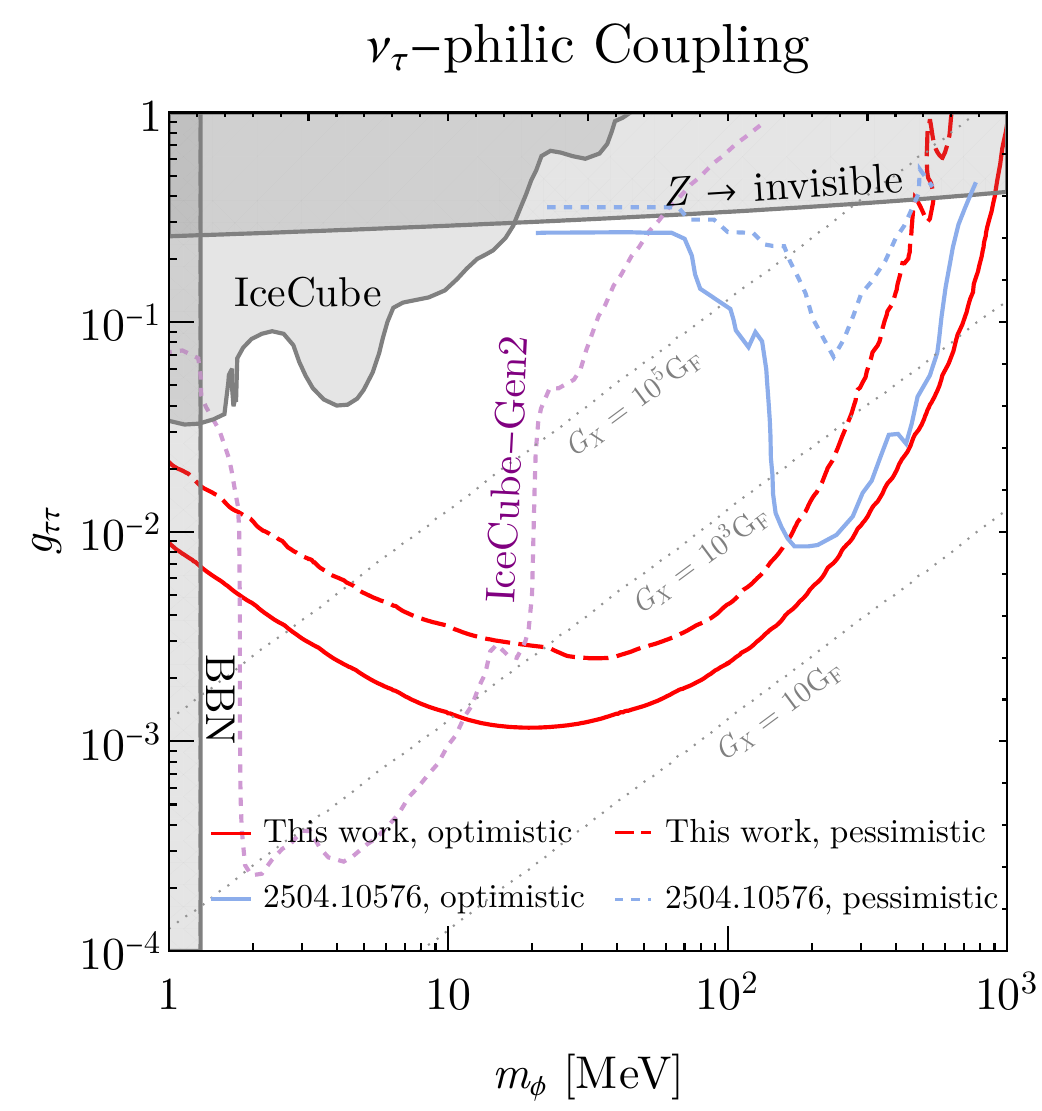}\includegraphics[width=0.49\textwidth]{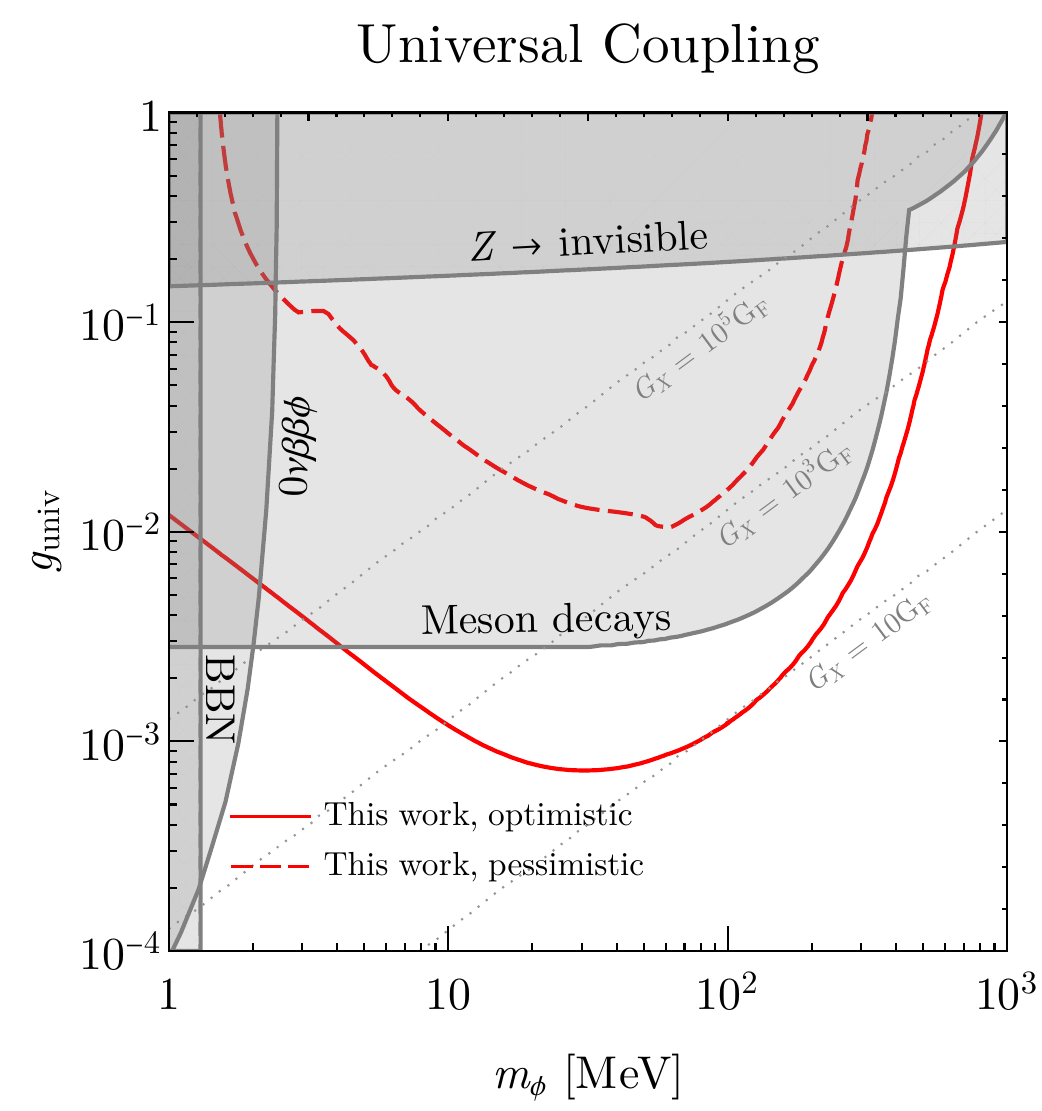}
\caption{\textbf{(Main result of the paper.)} Our projected sensitivities on $\nu$SI from UHE neutrinos absorbed by relativistic CNB, i.e., under the ``widening the resonance'' scenario, along with results from previous works.
\textbf{{\textcolor[HTML]{DC143C}{Red solid (dashed) curves:}}}
our sensitivities using the GRAND UHE neutrino detector, assuming optimistic (pessimistic) fluxes.
\textbf{{\textcolor[HTML]{2F80B9}{Blue curves:}}} previous results from UHE neutrinos absorbed by non-relativistic CNB with optimistic (solid) and pessimistic (dashed) fluxes~\cite{Leal:2025eou}.
\textbf{{\textcolor[HTML]{C080D9}{Purple dashed curve:}}}
earlier results from HE neutrinos absorbed by relativistic CNB and detected by IceCube-Gen2 (dashed)~\cite{Esteban:2021tub}.
\textbf{{\textcolor[HTML]{949494}{The gray shaded regions}}} are those excluded by $Z$ boson invisible decay~\cite{Brdar:2020nbj}, BBN~\cite{Blinov:2019gcj}, rare meson decay (universal coupling only)~\cite{Berryman:2018ogk}, neutrinoless double beta decay ($0\nu\beta\beta$), and IceCube HE neutrino observation ($\nu_\tau$-philic coupling only)~\cite{IceCube:2020wum}.
We note that our results depend on the assumption of the lightest neutrino mass being lighter than the CNB temperature, which is $T=1.9$~K or $m_3\lesssim 1.6\times 10^{-4}$~eV.
}
\label{fig:result}
\end{figure}

\section{Results}
\label{sec:results}

Figure~\ref{fig:result} shows our projected sensitivities (red curves) alongside experimental or projected constraints from earlier studies. 
The gray shaded regions represent known bounds from $Z$ boson invisible
decay~\cite{Brdar:2020nbj}, rare meson decays~\cite{Berryman:2018ogk}, majoron emission in neutrinoless double beta decay ($0\nu\beta\beta\phi$)~\cite{Brune:2018sab}, BBN~\cite{Blinov:2019gcj},
and IceCube HE neutrino observation~\cite{IceCube:2020wum}. Since $\nu_{\tau}$ is
not involved in meson decays and $0\nu\beta\beta\phi$, these bounds
do not apply to the $\nu_{\tau}$-philic scenario. 
Supernova constraints~\cite{Fiorillo:2022cdq} are not relevant in the shown parameter space, because $\phi$ has a lifetime that is too short
to escape the core of a supernova\footnote{This can be estimated as follows. According to Eq.~\eqref{eq:-13}, the mean free path of $\phi$ is roughly $16\pi E_{\phi}/(g^{2}m_{\phi}^{2})$,
which needs to be greater than the typical radius of a supernova core,
$R\sim10$ km. For $E_{\phi}$ taking the typical temperature of a supernova core, which is about $30$ MeV, $16\pi E_{\phi}/(g^{2}m_{\phi}^{2})\gtrsim10$
km leads to $g\lesssim10^{-7}\times(1\ \text{MeV}/m_{\phi})$, which
agrees well with the black dashed line in Fig. 1 of Ref.~\cite{Fiorillo:2022cdq}.}.
We note here that the lepton-number-violating feature of $\nu$SI may also cause novel effects in supernova such as neutrino-antineutrino and flavor equilibrium, as has been recently studied in Ref.~\cite{Suliga:2024nng}. Such effects have not yet been observable in current data but may be probed in future high-statistics observations of galactic SN explosions.

Our result demonstrates that future observations of UHE neutrinos by GRAND can probe $\nu$SI with the mediator mass up to 1 GeV and the coupling down to $10^{-3}$. 
For $\nu_\tau$-philic coupling, our result surpasses the existing limit from $Z$ boson invisible decay by two orders of magnitude.
For universal coupling, rare meson decays set stringent limits.
Our results, however, still surpass the current limits by a factor of a few for mediators heavier than 10 MeV. 

Finally, we emphasize the key difference between our work and earlier studies in Refs.~\cite{Esteban:2021tub,Leal:2025eou}. 
Thanks to the widened resonance arising from relativistic CNB, 
our sensitivities not only fill the gap between using HE~\cite{Esteban:2021tub} and UHE neutrinos~\cite{Leal:2025eou} scattering off the non-relativistic CNB, but also cover a significantly broader range of the mediator mass. 
Comparing with Ref.~\cite{Leal:2025eou}, which studies the case of non-relativistic CNB with the same UHE neutrinos and detection setup, we find that, if the lightest neutrino in the CNB is relativistic, the widened resonance can significantly expand the sensitivity reach of GRAND.
Reference~\cite{Esteban:2021tub} focused on TeV--PeV neutrinos, so the lower-mass regime of $\nu$SI was probed. Although not considered in this work, we anticipate that using the same TeV--PeV neutrinos and detection setup as Ref.~\cite{Esteban:2021tub}, if there is a relativistic component in the CNB,
the widened resonance would expand the sensitivity reach obtained in Ref.~\cite{Esteban:2021tub} as well.

\section{Conclusion}
\label{sec:conclusion}

Cosmic messengers play a central role in probing new physics beyond the standard model. This is because they provide experimental conditions far beyond the reach of laboratory experiments and serve as a link between laboratory discoveries of new fundamental physics and its role in the Universe, where many new physics motivations originate. On the other hand, neutrino self-interactions are well-motivated by nonzero neutrino masses, which typically require introducing new mediators that pass new forces between neutrinos. These $\nu$SI have rich phenomena in particle physics, astrophysics, and cosmology, and have been a vibrant subject in recent years.

In this work, we extend the 
``widening the resonance'' concept to ultrahigh-energy neutrinos and probe an uncharted region of parameter space of $\nu$SI in the high-mass regime.
The incoming UHE neutrinos scatter off the relativistic CNB due to $\nu$SI, leading to distortions of the UHE neutrino spectra.
The relativistic CNB exhibits a broad spectrum rather than the $\delta$-function-like spectrum of the non-relativistic CNB; this broadens the resonance of $\nu$SI and significantly enhances the spectral distortion. This feature requires that the lightest neutrino species is relativistic. Given that the CNB temperature $T=1.9$ K, the relativistic assumption corresponds to $m_{1}$  or $m_3\lesssim 1.6\times 10^{-4}$ eV for normal or inverted mass ordering, respectively.  Such an assumption is well compatible with neutrino oscillation data, well motivated in theories that predict one massless neutrino (e.g., the minimal seesaw model), and favored by recent DESI's limits on neutrino masses from cosmology~\cite{DESI:2024mwx}.

In addition to the feature of widened resonance, we also provide a semi-analytic framework to calculate the cosmogenic UHE neutrino production, avoiding computationally intensive simulations and precise enough for phenomenology studies.
This framework, which significantly simplifies the computation, can be directly applied to future phenomenology studies.

We solve the Boltzmann equations that govern the neutrino propagation, which reveal the spectrum distortion. 
Then, we use GRAND, a future powerful UHE neutrino detector, to probe such distorted spectra. Using proper likelihood analyses, we derive our projected sensitivities.
The widened resonance provides key advantages: it affects a wider range of UHE neutrino spectra, yielding greater statistics and overcoming the potential energy-resolution limitation.
Our sensitivities with optimistic UHE neutrino fluxes show that GRAND can probe the $\nu$SI mediator with MeV-GeV masses with couplings down to $g\sim 10^{-3}$, corresponding to $G_X\sim 10 G_{\rm F}$. 
Our sensitivities with pessimistic flux are weaker by a factor of 2--5 in the $\nu_\tau$-philic-coupling case and 10--50 in the universal-coupling case.
Overall, our sensitivities lead to improvements of up to two orders of magnitude over current constraints, as well as significant gains with respect to the non-relativistic CNB case.

Looking forward, continued experiments with larger effective areas in UHE neutrino detection and improved precision in modeling UHE neutrino production will make UHE neutrinos even sharper tools for probing fundamental physics.
The continuous detection of UHE neutrinos will either place our theoretical forecasts under experimental constraints or lead to discovery.

Our work illustrates a novel way of probing $\nu$SI.
The 
``widening the resonance'' is a general concept when both initial states follow continuum energy distributions. Thus, more phenomenology studies along this direction can be performed.

\section*{Acknowledgement}
P.M., I.R.W., and B.Z. are supported by Fermi Forward Discovery Group, LLC under Contract No. 89243024CSC000002 with the U.S. Department of Energy, Office of Science, Office of High Energy Physics.
I.R.W. is also supported by DOE distinguished scientist fellowship grant FNAL 22-33.
X.J.X is supported in part by the National Natural Science Foundation of China under grant No.~12141501 and also by the CAS Project for Young Scientists in Basic Research (YSBR-099).
We are grateful for the computing resources of Rutgers CMS group.

\appendix

\titleformat{\section}{\large\bfseries}{\thesection}{1em}{}
\section{More details for Section~\ref{sec:source}---modeling cosmogenic UHE neutrino production}

\label{app:source}

In this section, following Ref.~\cite{Kelner:2008ke}, we provide the necessary details for $\Psi_i(\eta, x)$ in Eq.~\eqref{eq:dNdEnu}, which is the neutrino spectrum per proton-photon interaction.

The scattering cross sections of photopion production are determined by the CoM energy, parametrized by $\eta$.
The kinematics gives a lower bound on $\eta$,
\begin{align}
\eta \geq \eta_0 \equiv 2 \frac{m_\pi}{m_p} + \frac{m_\pi^2}{m_p} \simeq 0.313\,.
\end{align}
Below this bound, $\Psi_i = 0$.
We further define $x_\pm$ related to the maximum and minimum energies of pions,
\begin{align}
    x_\pm \equiv \frac{1}{2(1+\eta)} \left[ \eta + r^2 \pm \sqrt{(\eta - r^2 - 2r)(\eta - r^2 + 2r)} \right]\,,
\end{align}
where $r \equiv m_\pi/m_p \simeq 0.146$.

Ref.~\cite{Kelner:2008ke} provides an analytic form for $\Psi_i(\eta, x)$, which is obtained by fitting their simulated data,
\begin{align}
    \label{eq:scattering flux}
    \Psi_i(\eta, x) = B_i \exp \left[- s_i \left( \frac{x}{x'_-} \right)^{\delta_i}\right] \times \left[ \ln \left( \frac{2}{1+y'^2} \right) \right]^{\psi_i}\,.
\end{align}
Here $i$ stands for the neutrino species.
$B_i$, $s_i$, and $\delta_i$ are all functions of $\eta$, whose values are provided in Tables II and III of Ref.~\cite{Kelner:2008ke}.
The $y'$ is defined as
\begin{align}
    y' \equiv \frac{x - x'_-}{x'_+ - x'_-}\,.
\end{align}
The $\psi_i$ and the relation between $x'_\pm$ and $x_\pm$ are given species-by-species.
We define $\rho \equiv \eta/\eta_0$ for convenience.
The results are summarized as follows, and we refer the readers to Sec. II.B of Ref.~\cite{Kelner:2008ke} for more detailed explanations.

{\bf\boldmath $\bar{\nu}_\mu$ and  $\nu_e$:}
\begin{align}
\psi = 2.5 + 1.4\ln \rho\,, ~~ x'_- = x_-/4\,,~~ x'_+ = x_+\,.
\end{align}

{\bf \boldmath$\nu_\mu$:}
\begin{align}
& \psi = 2.5 + 1.4\ln \left( \rho\right)\,,~~ x'_- = 0.427 x_-\,,\nonumber          \\
x'_+ & = \begin{cases}
0.427 x_+, \rho< 2.14\,,                                                        \\
x_+\left(0.427 + 0.0729 \left(\rho - 2.14\right)\right) \,, 2.14 < \rho < 10\,, \\
x_+\,, \rho > 10\,.
\end{cases}
\end{align}

{\boldmath\bf $\bar{\nu}_e$:} The form for this flavor is different from others, as $\pi^+$ cannot produce $\bar{\nu}_e$ via its decay so that an additional $\pi^-$ needs to be produced simultaneously.
This requires
\begin{align}
\eta > 4 r (1+r) \simeq 2.14 \eta_0\,.
\end{align}
The definition of $x_\pm$ for $\bar{\nu}_e$ is
\begin{align}
x_\pm \equiv \frac{1}{2(1+\eta)} \left( \eta - 2r \pm \sqrt{\eta (\eta-4r(1+r))} \right)\,.
\end{align}
The other quantities needed in Eq.~\eqref{eq:scattering flux} is given by
\begin{align}
\psi = 6 \left( 1 - e^{1.5 (4-\rho)} \right) \Theta(\rho-4)\,, ~~x'_- = x_-/2\,, ~~x'_+ = x_+\,,
\end{align}
where $\Theta$ is the Heaviside function.

\small{\bibliography{main}}

\end{document}